\documentclass{article}

\usepackage{arxiv}

\usepackage{amssymb}
\usepackage{bm}
\usepackage{soul}
\usepackage{color}
\usepackage{amsmath}
\usepackage{comment}
\usepackage{float}

\usepackage{upgreek}

\usepackage{subcaption}
\usepackage{graphicx}
\usepackage{multirow}
\usepackage{makecell}

\newcommand{\bd}{\bm{d}}
\newcommand{\bG}{\bm{G}}
\newcommand{\bF}{\bm{F}}
\newcommand{\bR}{\bm{R}}

\newcommand{\bK}{\bold{K}}

\title{FE-PINNs: finite-element-based physics-informed neural networks for surrogate modeling}

\author{
 Pranav Sunil \\
  Department of Materials Science and Engineering\\
  Rutgers University\\
  Piscataway, NJ 08854 \\
  \texttt{ps1170@soe.rutgers.edu} \\
   \And
 Ryan B. Sills \\
  Department of Materials Science and Engineering\\
  Rutgers University\\
  Piscataway, NJ 08854 \\
  \texttt{ryan.sills@rutgers.edu}
}

\graphicspath{{figures/}}
\begin{document}

\maketitle

\begin{abstract}
We present a method whereby the finite element method is used to train physics-informed neural networks that are suitable for surrogate modeling.
The method is based on a custom convolutional operation called stencil convolution which leverages the inverse isoparametric map of the finite element method. 
We demonstrate the performance of the method in several training and testing scenarios with linear boundary-value problems of varying geometries. 
The resulting neural networks show reasonable accuracy when tested on unseen geometries that are similar to those used for training.
Furthermore, when the number of training geometries is increased the testing errors systematically decrease, demonstrating that the neural networks learn how to generalize as the training set becomes larger.
Further extending the method to allow for variable boundary conditions, properties, and body forces will lead to a general-purpose surrogate modeling framework that can leverage existing finite element codes for training. 
\end{abstract}





\section{Introduction}
\label{sec:intro}

In the last few years, a major focus of the machine learning community has been on the use of neural networks (NNs) for surrogate modeling and solving partial differential equations (PDEs).
The basic idea is to replace a conventional modeling or simulation technique with a NN in order to reduce computational time.
These methods can be broken into two categories: data-driven NNs and physics-informed NNs.
Data-driven NNs are trained on the basis of labeled data, typically in the form of solutions from traditional simulation techniques such as the finite element (FE), finite volume, and finite difference methods~\cite{tripathyDeepUQLearning2018,kovachkiNeuralOperatorLearning2023}.
The training loss is the norm of the difference between the NN prediction and the labeled dataset.
This approach thus requires obtaining solutions in advance of training, which can be computationally burdensome.
Building on works from the late 1990s~\cite{lagarisArtificialNeuralNetworks1998}, Raissi et al.~\cite{raissiPhysicsinformedNeuralNetworks2019} introduced the physics-informed NN (PINN) method (similar to the method of Sirignano and Spiliopoulos~\cite{sirignanoDGMDeepLearning2018}), whereby the training loss function is the residual of the governing differential equation.
PINNs offer the major advantage that labeled data is not required, thereby accelerating their development.
Since Raissi et al.'s work, there has been a tremendous flurry of activity in applying PINNs to numerous problems including solid mechanics~\cite{haghighatPhysicsinformedDeepLearning2021,raoPhysicsInformedDeepLearning2021}, fluid mechanics~\cite{caiPhysicsinformedNeuralNetworks2021FM}, heat transfer~\cite{caiPhysicsInformedNeuralNetworks2021HT}, time-dependent problems~\cite{mengPPINNPararealPhysicsinformed2020}, and stochastic PDEs with uncertainty~\cite{nabianDeepLearningSolution2019,zhuPhysicsconstrainedDeepLearning2019,karumuriSimulatorfreeSolutionHighdimensional2020}.
PINNs have also been extended to variational loss functions rather than strong form PDEs by a number of authors~\cite{nabianDeepLearningSolution2019,kharazmiVariationalPhysicsInformedNeural2019,kharazmiHpVPINNsVariationalPhysicsinformed2021}.

One major drawback for many of the prior PINN works is that the geometry, boundary conditions (BCs), initial conditions, and properties cannot be changed after training.
In order to obtain a solution under different conditions, the PINN must be retrained. 
This makes it difficult to use PINNs as surrogate models, which can provide solutions for varying scenarios.
A number of researchers have extended PINN methods to enable variable conditions (e.g., BCs, geometry, properties) when evaluating the PINN.
Sun et al.~\cite{sunSurrogateModelingFluid2020} implemented fully connected (FC) PINNs with variable input parameters for specifying simulation parameters (e.g., fluid viscosity) and domain geometry.
Their method requires that the domain geometry can be described parametrically (e.g., with a mathematical function).
Hence, arbitrary geometries cannot be accommodated in their method.
Joshi et al.~\cite{joshiGenerativeModelsSolving2019} and Zhu et al.~\cite{zhuPhysicsconstrainedDeepLearning2019} developed convolutional neural network (CNN) based generative methods for obtaining PDE solutions with variable boundary conditions and/or parameters, utilizing finite differences to approximate derivatives on a uniform grid.
The requirement of a uniform grid is, of course, rather restrictive as it does not allow for local solution refinement and is difficult to simulate arbitrary domain geometries.
Gao et al.~\cite{gaoPhyGeoNetPhysicsinformedGeometryadaptive2021} developed the PhyGeoNet method which allows for irregular domains with variable properties and boundary conditions.
The core of their method is finite difference approximations coupled with a transformation mapping that maps from the physical domain with a non-uniform grid to a reference domain with a uniform grid.
This enables for arbitrary domain geometry as long as the transformation mapping can be established, which may not always be possible.
Additionally, with PhyGeoNet it is difficult to refine the discretization locally.

BC enforcement is another challenging aspect of PINNs.
The traditional approach is to include an additional term in the loss function which captures deviation of the NN solution from the chosen BCs~\cite{sirignanoDGMDeepLearning2018,raissiPhysicsinformedNeuralNetworks2019}, which is often referred to as soft BC enforcement since the BCs are not exactly enforced.
With soft enforcement, there is a competition during training between accuracy of the solution within the domain interior versus on the boundaries.
To overcome this issue, a number of authors have developed hard BC enforcement techniques which exactly satisfy the BCs.
One approach is to employ a distance function which smoothly ``turns off'' the NN solution at the boundaries, so that the appropriate BC values can be manually added to the solution~\cite{sunSurrogateModelingFluid2020}.
Specifying this distance function for arbitrary domains is not trivial, however, with several researchers employing pre-trained NNs to serve as the distance function~\cite{sunSurrogateModelingFluid2020,shengPFNNPenaltyfreeNeural2021}.
In the PhyGeoNet technique of Gao et al.~\cite{gaoPhyGeoNetPhysicsinformedGeometryadaptive2021}, hard BC enforcement is accomplished by simply manually specifying the solution at boundary grid points.

A final challenge for PINNs is the use of CNNs, which are more difficult to implement in PINNs compared to FC-NNs.
It is well established that CNNs exhibit superior scaling and trainability in comparison to FC-NNs owing to their parameter sharing~\cite{Goodfellow-et-al-2016}.
Furthermore, with FC-NNs the problem size (e.g., number of grid points) cannot easily be changed, since the number of inputs is fixed at training.
CNNs do not have this limitation, since convolutions can be performed for any input size and shape.
However, the challenge in employing CNNs is that a uniform grid is required~\cite{joshiGenerativeModelsSolving2019,zhuPhysicsconstrainedDeepLearning2019}, which limits the domain geometry as discussed above.
Gao et al.~\cite{gaoPhyGeoNetPhysicsinformedGeometryadaptive2021} overcame this issue by performing convolutional operations in the uniform reference domain rather than the non-uniform physical domain.
An alternative method is to employ a graph neural network (GNN), which performs convolutions using a graph comprised of nodes connected by edges rather than a grid.
If this graph represents a computational mesh defining the simulation domain (e.g., a finite element mesh), then the geometry can be arbitrary.
Gao et al.~\cite{gaoPhysicsinformedGraphNeural2022} recently implemented this approach using a variational PINN.
While powerful, GNNs still impose restrictions on the discretization, since every node of the graph must have the same connectivity.
Furthermore, graph convolutions have the drawback that the convolutional environment for each node is different if the mesh is unstructured.
This is likely to make GNN-based PINNs sensitive to the computational mesh.

Our goal in this work is to overcome these three key challenges (variable and arbitrary geometric domains, strong enforcement of BCs, use of CNNs in PINNs) by coupling a physics-informed CNN to the finite element (FE) method.
We call this approach the FE-based PINN (FE-PINN) method.
It is conceptually similar to the variational GNN method of Gao et al.~\cite{gaoPhysicsinformedGraphNeural2022}, but differs through the introduction of a new convolutional operation which we call stencil convolution.
The stencil convolution is very similar to a traditional convolution, and so does not suffer the drawbacks of graph convolutions discussed above.
Stencil convolution makes use of the FE approximation of the solution field, making the FE-PINN method similar to the neural operator methodology developed by Kovachki et al.~\cite{kovachkiNeuralOperatorLearning2023} since the input and output for a FE-PINN are piece-wise functions (derived from FE basis functions).
Finally, FE-PINNs utilize the standard FE method to compute the physics-informed loss function, making the FE-PINN methodology readily extensible to any physical problem which is currently solved via the FE method.

In the remainder of the manuscript, we will first review the relevant basics of the FE method.
We will then present the FE-PINN method, with the main novelty being the introduction of stencil convolution. 
Next we will present the performance of FE-PINNs for two different classes of geometries where the training and testing geometries are varied.
Finally we discuss potential use cases and future research topics for FE-PINNs.

\section{Methods}
\label{sec:methods}

\subsection{Basics of the FE method}
Here we briefly review the FE method with emphasis on its features of relevance to FE-PINNs.
The FE method is based on the so-called weak form of an initial-boundary value problem.
This form can be obtained from the strong form, which typically manifests as a PDE (e.g., the heat equation, the equations of equilibrium, the Navier-Stokes equations), by multiplying with a so-called weight function and applying integration by parts~\cite{hughesFiniteElementMethod2000}. 
As an example which is the focus of our work here, consider the equations of equilibrium in strong form:
\begin{eqnarray}
    \frac{\partial\sigma_{ij}(u_k)}{\partial x_j} + f_i &=& 0 \quad \text{in $B$} \\
    u_i &=& g_i \quad \text{on $\partial B_{u_i}$} \\
    \sigma_{ij}n_j &=& h_i \quad \text{on $\partial B_{t_i}$}
    \label{eq:equil_strong}
\end{eqnarray}
where $x_j$ is the $j$-th coordinate direction, $u_i$ is the displacement in the $i$-th coordinate direction, $B$ is the interior of the body, $\partial B_{u_i}$ and $\partial B_{t_i}$ are portions of the boundary where displacement BCs $g_i$ and traction BCs $h_i$ are applied, respectively,  $\sigma_{ij}$ is the stress tensor, $f_i$ is the body force, and repeated indices imply summation.
Applying the procedure discussed above, one obtains the weak form of the equations of equilibrium as~\cite{hughesFiniteElementMethod2000}:
\begin{equation}
    \int_B \frac{\partial w_i}{\partial x_j} \sigma_{ij}(u_k) dV - \int_B w_i f_i dV \, - \, \int_{\partial B_{t_i}} w_i h_i dA = 0
    \label{eq:equil_weak}
\end{equation}
where $w_i$ is the so-called weight function for the $i$-th coordinate direction.
The set of possible displacement fields $u_i$ is constrained to automatically satisfy the displacement BCs, while the set of possible weight functions $w_i$ is constrained to equal zero on $\partial B_{u_i}$.

To obtain the weak Galerkin form employed by the FE method, one limits the set of possible displacement and weight functions to a user-selected set.
These functions are made up of so-called shape functions, which will be discussed further below.
Let $v_i$ be a function taken from the space of allowable shape functions which also satisfies the constraint that $v_i = 0$ on $\partial B_{u_i}$.
We can thus express the displacement field as $u_i = v_i + g_i$ and automatically enforce the displacement BCs.
Using this expansion of $u_i$, we can re-express Eq.~(\ref{eq:equil_weak}) as 
\begin{equation}
    \int_B \frac{\partial w_i}{\partial x_j} \sigma_{ij}(v_k) dV - \int_B w_i f_i dV \, - \, \int_{\partial B_{t_i}} w_i h_i dA + \int_B \frac{\partial w_i}{\partial x_j} \sigma_{ij}(g_k) dV = 0
    \label{eq:equil_Gweak}
\end{equation}
Eq.~(\ref{eq:equil_Gweak}) is the statement of the problem used in the FE method.
Solving it requires determining the function $v_i$ which satisfies the equality for all functions $w_i$ within the allowable set.

To proceed with solving Eq.~(\ref{eq:equil_Gweak}) a procedure must be established for evaluating each of the integrals.
To accomplish this, in the FE method the problem domain $B$ is discretized into a set of non-overlapping finite elements, which are polygons/polyhedra of user-defined geometry (e.g., triangles or quadrilaterals for 2D problems, tetrahedra or hexahedra for 3D problems).
The geometry of each element $e$ is defined by a set of nodes and each node $n$ is associated with a so-called elemental shape function, $\tilde{N}^{ne}(x_i)$ (the tilde denotes evaluation in the physical domain, see discussion below).
These shape functions are used to discretize the solution field $u_i(x_i)$ as a weighted sum over the shape functions:
\begin{equation}
    u_i(x_i;\bm{d})=\sum_e\sum_{n\in e} d_i^n \tilde{N}^{ne}(x_i) 
    \label{eq:u}
\end{equation}
where $d_i^n$ is the displacement value for coordinate direction $i$ at node $n$.
The solution field is then fully specified by the nodal values $d_i^n$, which are collected together into a global solution vector $\bd$.

The elemental shape functions are designed to have compact support, meaning that they are only nonzero within their respective elements.
Furthermore they are defined in the so-called parent domain of each element, where the element has a regular geometry.
In contrast, in the physical domain of the problem each finite element will, in general, have an irregular geometry.
The mapping that connects the two domains is the so-called isoparametric mapping~\cite{hughesFiniteElementMethod2000}, given by:
\begin{equation}
    x_i(\xi_i)=\sum_e\sum_{n\in e} X_i^n N^{ne}(\xi_i) 
    \label{eq:isomap}
\end{equation}
where $X^n_i$ is the $i$-th coordinate of node $n$ in the physical domain and $N^{ne}(\xi_i)$ is the elemental shape function evaluated at the parent domain coordinate $\xi_i$.
Note that with these definitions $\tilde{N}^{ne}(x_i)=N^{ne}(\xi_i(x_i))$, where $\xi_i(x_i)$ is the inverse isoparametric mapping.
We emphasize that this mapping from parent to physical domain can be applied to arbitrary geometries (as long as elements are not excessively distorted~\cite{hughesFiniteElementMethod2000}), in contrast to the mapping methodology employed by Gao et al.~\cite{gaoPhyGeoNetPhysicsinformedGeometryadaptive2021} in PhyGeoNet.
As an example, for a bilinear quadrilateral element in 2D (used for all results presented below) the shape functions are:
\begin{eqnarray}
    N^{(1)}(\xi_i) &=& \frac{1}{4}(1-\xi)(1-\eta) \nonumber \\
    N^{(2)}(\xi_i) &=& \frac{1}{4}(1+\xi)(1-\eta) \nonumber \\
    N^{(3)}(\xi_i) &=& \frac{1}{4}(1+\xi)(1+\eta) \nonumber \\
    N^{(4)}(\xi_i) &=& \frac{1}{4}(1-\xi)(1+\eta)
    \label{eq:bilin_sf}
\end{eqnarray}
where $(\xi,\eta)\in[-1,1]$ are the parent coordinates and the notation $N^{(l)}$ means $l$ is the local node number for the element rather than the global node number.
The fact that each shape function has compact support means that the integrals in Eq.~(\ref{eq:equil_Gweak}) can be decomposed into a set of subintegrals, with each subintegral confined to a finite element.
Subintegrals are evaluated numerically using Gaussian quadrature in the parent domain.


Applying the discretization in Eq.~(\ref{eq:u}) to the displacement field and weight functions (thereby constraining the space of possible functions) in Eq.~(\ref{eq:equil_Gweak}) and then evaluating each of the elemental subintegrals via Gaussian quadrature, one arrives at the FE governing equations, which may be generally abstracted as~\cite{wriggersNonlinearFiniteElement2008}
\begin{equation}
    \bG(\bd) = \bm{0}.
    \label{eq:G}
\end{equation}
$\bG(\cdot)$ is a (possibly nonlinear) algebraic operator that captures (1) the governing equations within the domain that give rise to an ``internal force vector'' $\bF_{\rm i}(\cdot)$ (first term of Eq.~(\ref{eq:equil_Gweak})) and (2) boundary conditions/body forces/field sources which contribute to an ``external force vector'' $\bF_{\rm e}(\cdot)$ (other terms in Eq.~(\ref{eq:equil_Gweak})).
One can thus recast $\bG(\cdot)$ as a ``force balance'' between internal and external forces, $\bG(\cdot) = \bF_{\rm i}(\cdot)-\bF_{\rm e}(\cdot)$.
Note that the terminology of ``force'' here is a historical one and does not imply that the FE method is limited to mechanics problems. 
On the contrary, it is commonly applied to heat/mass transfer, fluid dynamics, and electromagnetism.

In the case of a linear problem (e.g., Fourier thermal conduction, small-deformation linear elasticity), Eq.~(\ref{eq:G}) may be rewritten as
\begin{equation}
    \bG(\bd) = \bK\cdot\bd - \bF_{\rm e} = \bm{0}.
    \label{eq:Kd-F}
\end{equation}
where $\bK$ is the so-called global stiffness matrix.
Solving a linear finite element problem thus simplifies to solving a linear system for the global solution vector $\bm{d}$.
$\bK$ is always a sparse matrix due to the compact support of the elemental shape functions.


In the FE method, BCs are termed as either being essential or natural.
Essential BCs correspond to direct specification of the field values (e.g., displacements), whereas natural BCs correspond to specification of a flux or traction.
An important feature of the FE method in comparison to other numerical methods and traditional methods for computing PINN residuals is that the influence of BCs is directly incorporated into the governing weak Galerkin form, Eq.~(\ref{eq:G}), through the external force vector.
Solving the equations automatically enforces the BCs.
This means that during training no additional action needs to be taken to enforce the BCs; if the FE-PINN trains well such that the equation residuals are small, the BCs are automatically (weakly) satisfied.

\subsection{FE-PINN methodology}
The FE-PINN method is based on three features: (1) the weak Galerkin form is used to compute the physics loss of the NN, (2) BCs are enforced using standard FE methods, and (3) convolutional operations are performed directly on the FE mesh using stencil convolutions.
We now discuss these features in detail.

\subsubsection{Loss computation and BC enforcement}
Based on the discussion above, the residual vector of the weak Galerkin form given some approximate solution $\tilde{\bd}$ (e.g., obtained from a surrogate model) can be expressed as
\begin{equation}
    \bR(\tilde{\bd}) \equiv \bG(\tilde{\bd}).
\end{equation}
We can use this residual to define a physics-informed loss function as
\begin{equation}
    \mathcal{L}(\tilde{\bd}) = \Vert \bR(\tilde{\bd}) \Vert
    \label{eq:loss}
\end{equation}
where in this case $\tilde{\bd}$ is the solution estimate given by the NN.
The benefit of formulating the loss function in this way is that it is readily computable using existing FE codes which already compute equation residuals.
This means that, in principle, a user can train a FE-PINN for any physical problem and any constitutive law which is implemented in the existing FE code.
Another advantage of formulating the loss function in this way is that it automatically enforces the BCs.
As discussed previously, in the Galerkin weak form BCs appear as additional ``force'' contributions in the governing equations, and thus contribute directly to the loss function.
No additional steps need to be taken when training the NN to enforce BCs and no arbitrary parameters need to be specified in the loss function to decide the relative importance of BC and residual losses.

\subsubsection{Stencil convolution}

The main contribution of our method is the introduction of stencil convolution.
The major challenge of employing a CNN with an FE problem is the irregularity of the mesh, which prohibits traditional, grid-based convolutional operations.
One solution, as pursued by Gao et al.~\cite{gaoPhysicsinformedGraphNeural2022}, is to use graph convolutions. 
However, graph convolutions are strongly influenced by the mesh and lack flexibility in their definition.
Here we propose a new type of convolution called \emph{stencil convolution} which emulates a traditional grid-based convolution while directly leveraging the approximations of the FE method.
Below we illustrate the stencil convolution method in two-dimensions, but it readily generalizes to three-dimensions.

\begin{figure}
    \centering
    \begin{tabular}{cc}
        \includegraphics[width=0.13\textwidth]{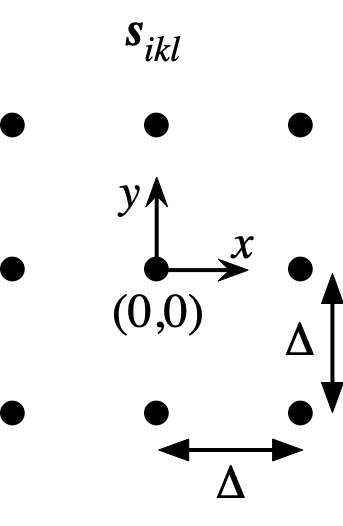} & 
        \includegraphics[width=0.66\textwidth]{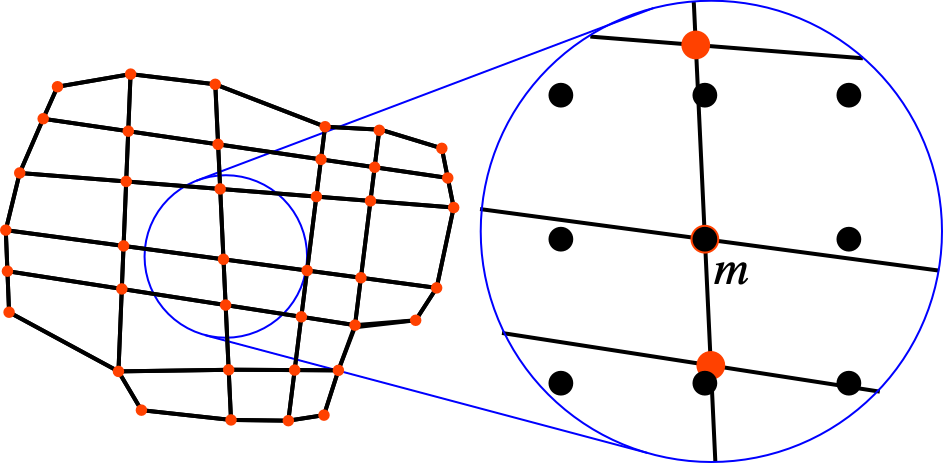}  \\
        (a) & (b) 
    \end{tabular}
    \caption{Basics of stencil convolution. (a) Stencil tensor definition and (b) stencil convolution performed on node $m$ of an FE mesh.}
    \label{fig:stencil}
\end{figure}

The main idea is to introduce a stencil of points, defined by a \emph{stencil tensor} $s_{ikl}$, which represents the local neighborhood around each node in the FE mesh that is to be used for convolution, as shown in Fig.~\ref{fig:stencil}(a).
For example, the following stencil tensor gives a uniform grid of spacing $\Delta$:
\begin{equation}
    s_{ikl}=
    \begin{cases}
    k\Delta, & i=0 \\
    l\Delta, & i=1
    \end{cases}
    \label{eq:sikl}
\end{equation}
 where $i$ denotes the coordinate direction, $k,l\in[-\omega,\omega]$ are integers indicating the stencil point, $\omega=(w-1)/2$, and $w$ is the odd, integer convolutional kernel size.
As the NN evaluates within the hidden layers, as long as the result is always stored on each node (just like the final solution vector), the FE approximation Eq.~(\ref{eq:u}) can be used to determine the value of the (hidden) field at any spatial point.
For each node, we evaluate the field at each stencil point, as shown in Fig.~\ref{fig:stencil}(b).

Let $\hm{h}_t$ denote the global vector of nodal values of the field for input channel $t$ at the current hidden layer/neuron of the NN.
The values of the field at the stencil points for node $m$ are
\begin{equation}
    \hat{I}_{tklm} = I(x^m_{ikl};\bm{h}_t)
\end{equation}
where $x^m_{ikl} \equiv X_i^m\bar{\delta}_{kk}\bar{\delta}_{ll}+s_{ikl}$ are the positions of the stencil points, $\bar{\delta}_{kk}= \delta_{kk}/2$, $\delta_{ij}$ is the Kronecker delta, and $I(\cdot)$ is the input field obtained using Eq.~(\ref{eq:u}).
The stencil convolution can then be computed as follows:
\begin{equation}
    O_{rm} = \sum_{t=1}^{N_{\rm in}} \sum_{k=-\omega}^{\omega}\, \sum_{l=-\omega}^{\omega} W_{rtkl} \cdot \hat{I}_{tklm} + b_r\bar{\delta}_{mm}
    \label{eq:stencil_conv}
\end{equation} 
where $O_{rm}$ is the output for output channel $r$ at node $m$, $W_{rtkl}$ is the tensor of convolutional weights, $N_{\rm in}$ is the number of input channels, and $b_r$ is the bias for output channel $r$.
Given a stencil tensor, this operation can be performed at each node, the output of which defines a new field which can be used as input to the next convolutional layer of the NN.

In order for stencil convolutions to be employed at scale, they must be efficiently implemented.
This can be accomplished as follows.
To determine the value of the input field $I$ at the stencil points $x^m_{ikl}$, we must determine the value of the elemental shape functions at those points.
Only shape functions for the element containing each point will be non-zero, thanks to their compact support. 
Thus, for each node we must determine which element the stencil points fall within.
Let $e^m_{(k,l)}$ be the element containing stencil point $(k,l)$ for node $m$.
Shape functions can only be explicitly evaluated in the parent domain (c.f. Eq.~(\ref{eq:bilin_sf})), so we must map from the physical stencil point coordinate $x^m_{ikl}$ to the parent coordinate $\xi^m_{ikl}$.
This is the inverse of the isoparametric mapping given in Eq.~(\ref{eq:isomap}).
For the bilinear quadrilateral elements used here the inverse isoparametric map can be evaluated analytically~\cite{huaInverseTransformationQuadrilateral1990}.

After identifying elements $e^m_{(k,l)}$ for each node and stencil point and evaluating the associated shape functions via the inverse isoparametric map, we can compute the stencil point values as
\begin{equation}
    \hat{I}_{tklm} = S_{klmn} \cdot h_{tn}
    \label{eq:Ihat}
\end{equation}
where $h_{tn}$ is the input (hidden) field value at node $n$ for input channel $t$ and $S_{klmn}$ is called the \emph{stencil convolution tensor}. 
The elements of $S_{klmn}$ are computed as
\begin{equation}
    S_{klmn} = 
    \begin{cases}
        N^{ne}(\xi^m_{ikl}(x^m_{ikl})) & \text{if}\,\, n \in e, e=e^m_{(k,l)} \\
        0 & \text{else}
    \end{cases}
    \label{eq:S}
\end{equation}
In other words, the stencil convolution tensor $S_{klmn}$ collects together all relevant shape function values so that the summation in Eq.~(\ref{eq:u}) can be efficiently computed as a tensor contraction.
Note that since the shape functions have compact support, $S_{klmn}$ will be a very sparse tensor.
$S_{klmn}$ is pre-computed prior to NN training since the mesh and stencil points do not change during training.
Once the tensor $\hat{I}_{tklm}$ is computed using Eq.~(\ref{eq:Ihat}), the convolution Eq.~(\ref{eq:stencil_conv}) can be evaluated as an additional tensor contraction.

\subsection{Neural network implementation, training, and testing}

Stencil convolution is a general class of NN operators which can operate on any input field which is defined on an FE mesh.
Hence, we could pass in fields indicating to the NN many details of the problem, e.g., geometry, BCs, properties (similar in nature to the PhyGeoNet approach~\cite{gaoPhyGeoNetPhysicsinformedGeometryadaptive2021}).
In this work we only present results where geometry is varied and plan to explore other problem variables in future work.
Hence, the input to our NNs at the input layer is the position vectors of the nodes; the number of input channels is 2, one for the nodal $x$-coordinates and one for the nodal $y$-coordinates.
FE-PINNs were trained on between one and three different geometries which comprised the training batch.
In the case of multiple training geometries, if the number of nodes differed between them they were zero-padded as appropriate to enable batched tensor operations.

\begin{figure}
    \centering
    \includegraphics[width=0.85\textwidth]{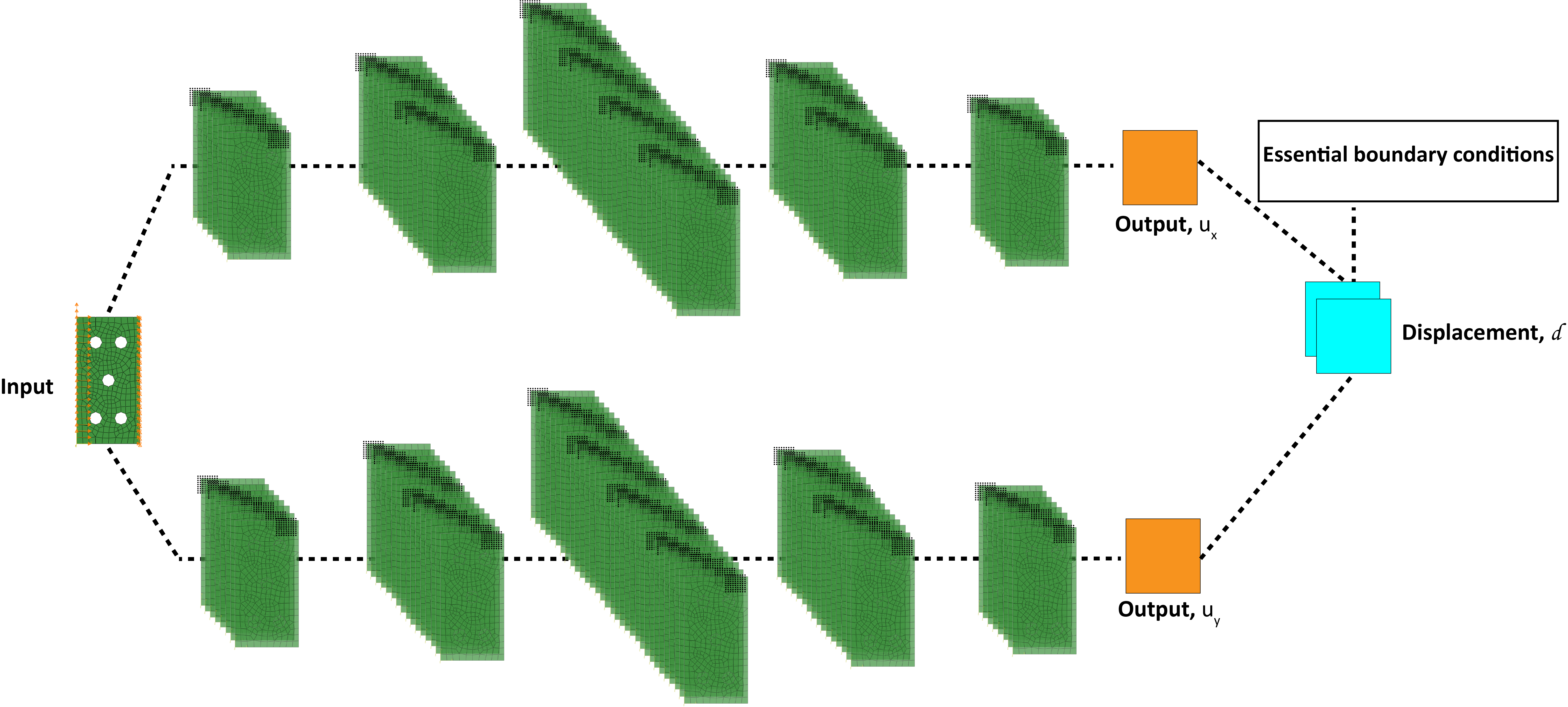} \\
        (a) \\
        \includegraphics[width=0.85\textwidth]{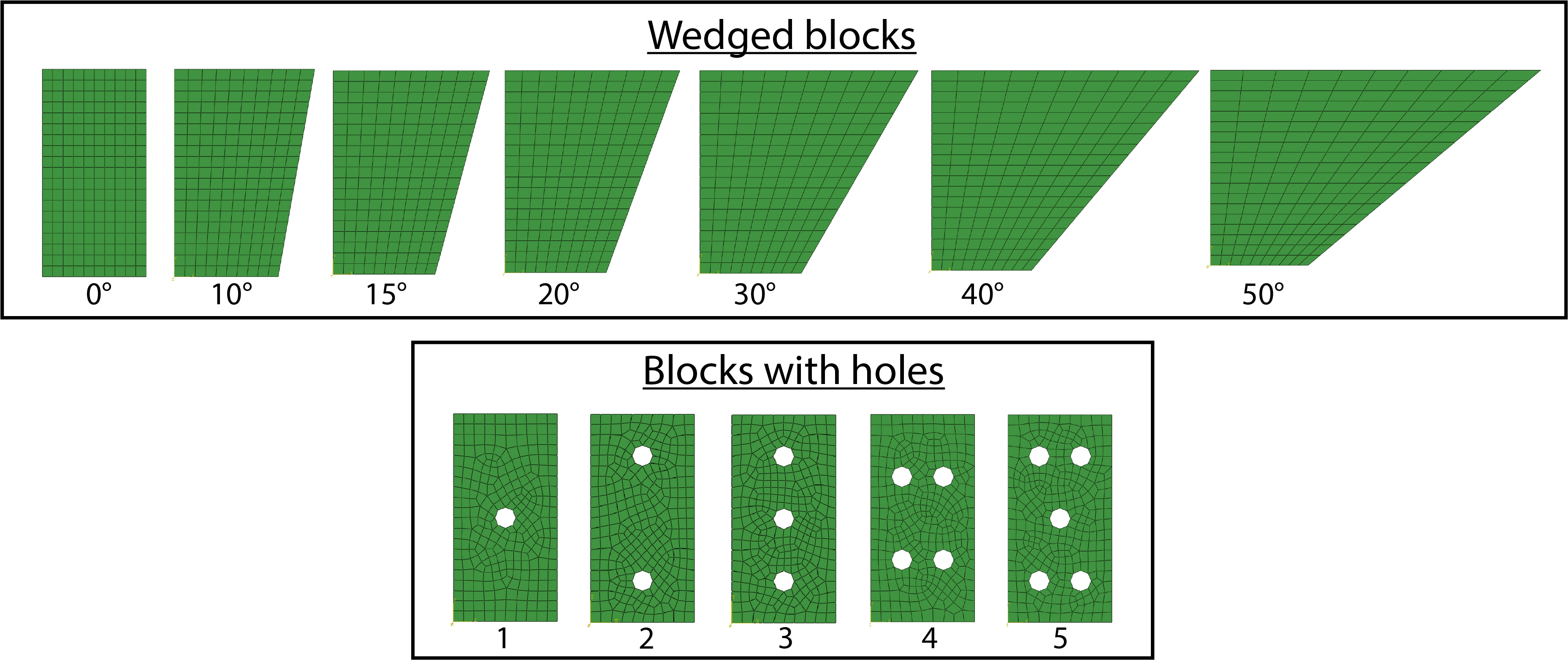}  \\
        (b) 
        
    \caption{(a) NN architecture used in this work. (b) FE geometries used for NN training and testing.}
    \label{fig:NN}
\end{figure}

All NN training was performed using PyTorch.
Our NN architecture is shown in Fig.~\ref{fig:NN}(a) and is similar to that used by Gao et al.~\cite{gaoPhyGeoNetPhysicsinformedGeometryadaptive2021}.
Rather than using a single NN to predict the $x$ and $y$ displacements, $u_x$ and $u_y$, we implemented separate sub-networks for each.
Each sub-net featured an input layer, three hidden layers, and an output layer.
The number of channels across the layers varied as follows: $2\rightarrow8\rightarrow16\rightarrow32\rightarrow16\rightarrow8\rightarrow1$.
Each hidden layer featured stencil convolutions followed by ReLU activation functions.
A convolutional kernel size of $w=9$ and the same stencil spacing $\Delta$ was used in all hidden layers.
Performance was tested with different values of $\Delta$.
The Adam optimizer was used with a learning rate of 0.001.
Training was performed for 1000 epochs, which was found to yield well-converged NNs.
For each training scenario, three different training runs were performed using different seeds to randomly initialize the weights and results are only presented for the best performer among the three.
Training times for all FE-PINNs was on the order of 1-minute using a single CPU.

After training on a set of chosen geometries, FE-PINNs were tested on a set of unseen geometries (see next section for more details). 
All losses presented below represent physics losses obtained using Eq.~(\ref{eq:loss}).
We have non-dimensionalized the presented losses (see below).

Stencil convolution was implemeneted as a custom NN operator in PyTorch.
The operator subroutine utilizes sparse storage of the stencil convolution tensor as a matrix by contracting together indices $(k,l,m)$, so that Eq.~(\ref{eq:Ihat}) can be expressed as a  sparse matrix-dense matrix product, yielding a dense matrix.
This is necessary to optimize computational performance and memory efficiency.
Eq.~(\ref{eq:stencil_conv}) is then evaluated using the {\tt einsum} operator in PyTorch.
To efficiently construct the stencil convolution tensor $S_{klmn}$, we used the GeoPandas package~\cite{kelsey_jordahl_2020_3946761} to determine which element each stencil point falls inside of via an {\tt sjoin} operation.
All backward gradients used during backpropagation were computed by PyTorch's autograd capability (no custom backward implementation was necessary).

\subsection{Finite element test cases}

The focus of the manuscript is on the training and testing of FE-PINN surrogate models for a chosen class of solid mechanics problems.
Two sets of geometries were used: a set of wedged blocks with variable wedge angle and a set of blocks with variable number of holes.
These geometries are shown in Fig.~\ref{fig:NN}(b), each has a height $H$ and the average element size in each mesh is $0.05H$.
Each of these sets of geometries provides a distinct set of characteristics for probing NN performance.
The wedged block geometries provide a smoothly graded variation in geometry, with each geometry being similar to nearby wedge angles.
However the nature of the loading varies as the wedge angle is changed since the BCs are unchanged.
On the other hand, the blocks with holes have the same external geometry but the internal geometry varies significantly; adding/removing holes represents a discontinuous change in geometry, rather than a smoothly graded one.
In all cases, the blocks are subjected to a fixed BC $u_x = u_y = 0$ on the right edge and unit displacements $u_x = u_y = 1$ on the opposing (left) edge.
The other two lateral edges were traction-free.
We used an isotropic, linear elastic material model for all cases with Young's modulus $E$ and Poisson's ratio $\nu=0.3$.
We also assume infinitesimal deformations in our solutions, rendering the problems linear (e.g., no geometric nonlinearity).
Hence, to compute our physics losses we made use of Eq.~(\ref{eq:Kd-F}) in our PyTorch code.
Stiffness matrices for each geometry were generated prior to training using ABAQUS 2019.
External force vectors resulting from the essential BCs were computed in our NN code using standard FE methods~\cite{hughesFiniteElementMethod2000}.
Meshes were generated using the ABAQUS mesh generation tool or a custom MATLAB script.
Plane strain, bilinear quadrilateral elements were used with 2x2 Gaussian quadrature (``fully integrated'' elements in ABAQUS terminology).

\begin{figure}
    \centering
    \begin{tabular}{cc}
        \includegraphics[width=0.47\textwidth]{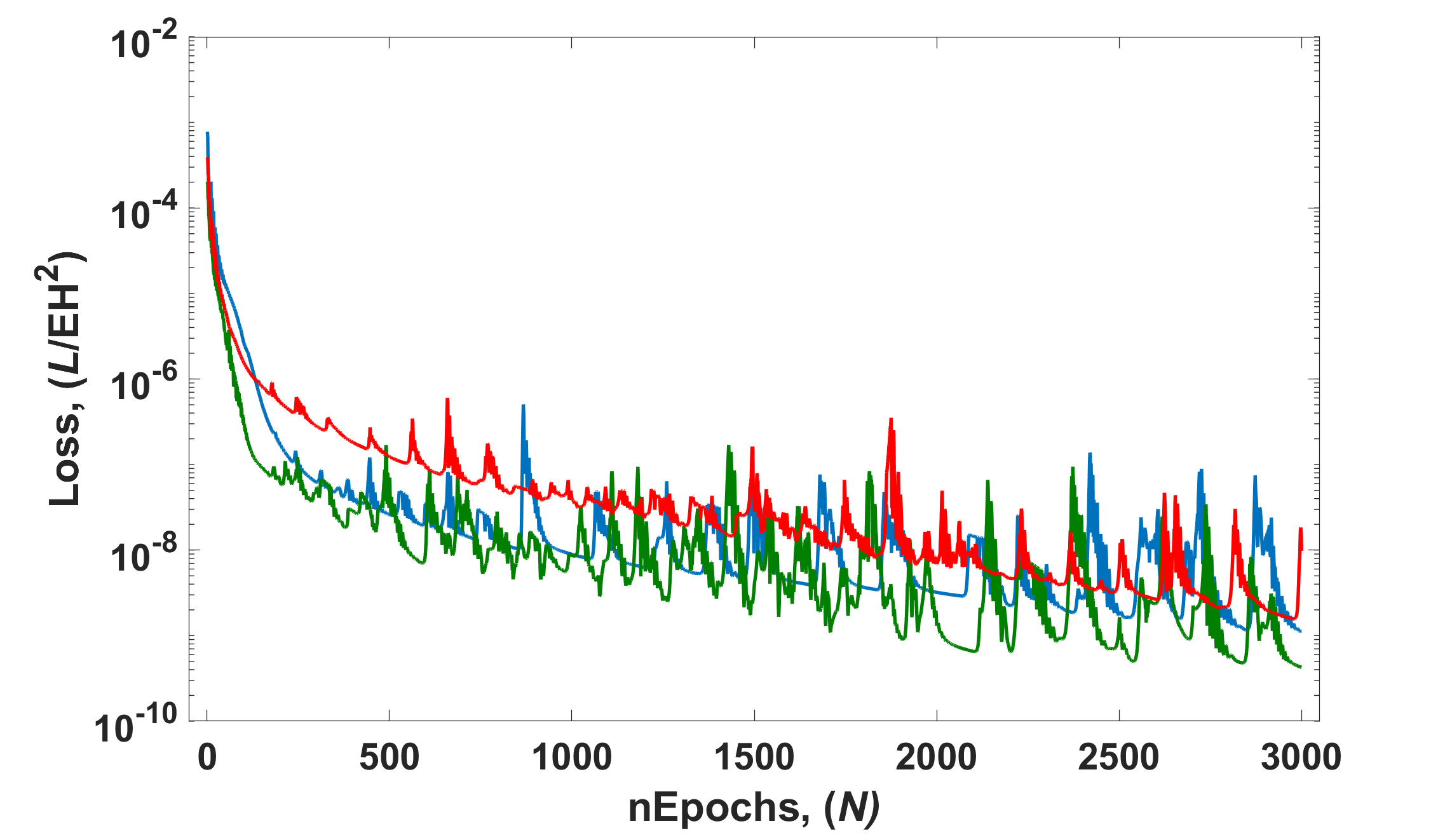} & 
        \includegraphics[width=0.47\textwidth]{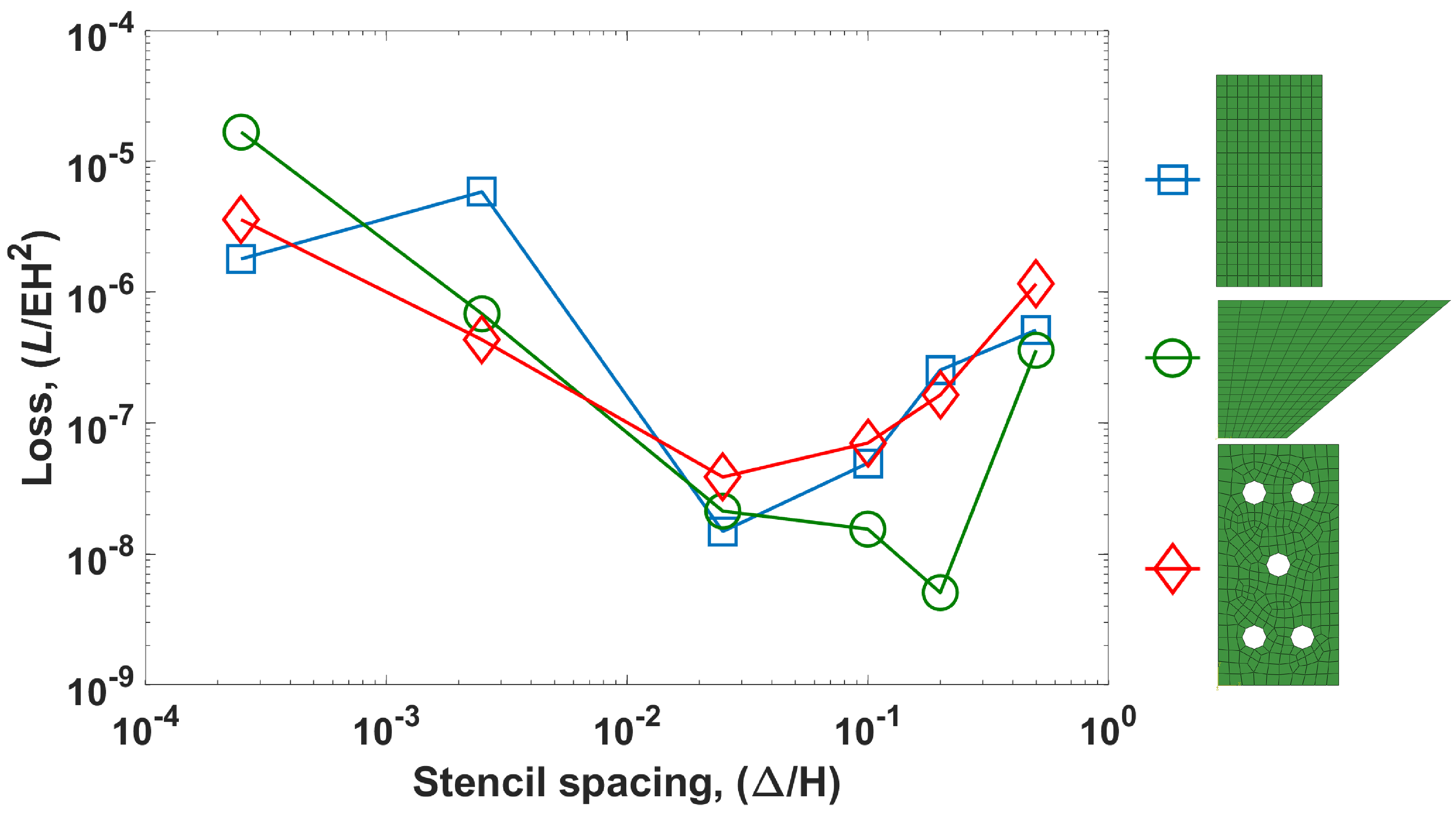}  \\
        (a) & (b) 
    \end{tabular} \\
    \includegraphics[width=0.95\textwidth]{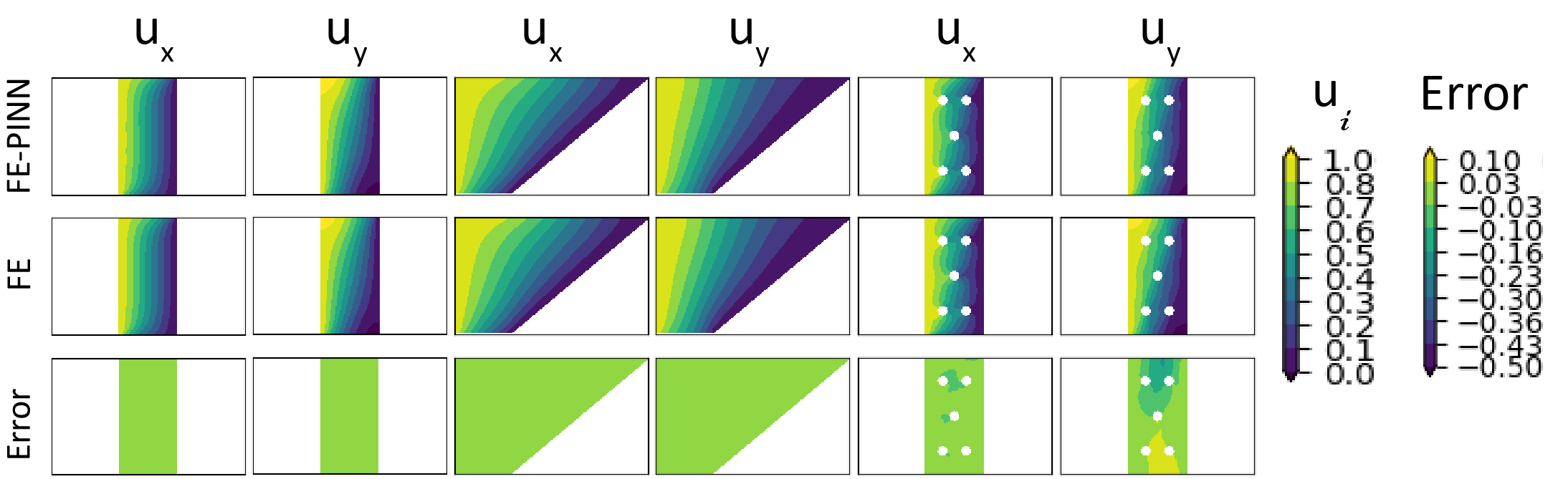} \\
    (c)
    \caption{Training convergence for three different geometries. (a) Training loss versus training epoch number. (b) Loss after 1000 epochs as a function of the stencil spacing $\Delta$. (c) Comparison of FE-PINN (top row) and FE (middle row) displacement fields, and difference between them (bottom row).}
    \label{fig:convergence}
\end{figure}

\section{Results}
\label{sec:results}

\subsection{Training convergence}

We begin by demonstrating the training convergence properties of FE-PINNs.
Fig.~\ref{fig:convergence}(a) shows training convergence curves for the three different geometries, plotting loss as a function of epoch number.
Physics losses computed using Eq.~(\ref{eq:Kd-F}) are presented non-dimensionally.
This figure shows robust convergence for all three geometries.
Note that the three geometries span different mesh categories: the unwedged block geometry has a structured and undistorted mesh, the wedged block geometry has a structured but distorted mesh, and the block with 5 holes geometry has an unstructured (and distorted) mesh.
NN convergence is shown to be insensitive to the nature of the mesh and is well-converged by 1000 epochs.
Going forward, all results are presented using NNs trained over 1000 epochs.

Fig.~\ref{fig:convergence}(b) shows the final loss (after 1000 epochs) as a function of the stencil spacing $\Delta$ used during the stencil convolutions.
For all three geometries there is a clear ``sweet spot'' where the training loss is lower.
If the stencil spacing is too small or too large, the field values at the stencil points are less informative to the NN leading to inferior training convergence.
Going forward, all FE-PINNs are trained with a stencil spacing of $\Delta/H=0.025$.

Finally, to visualize the quality of the FE-PINN solutions during training, we present in Fig.~\ref{fig:convergence}(c) displacement fields $u_x$ and $u_y$ for the three geometries obtained from the FE-PINN (top row), the FE solution (middle row), and the difference between the two (bottom row).
The displacement fields are obtained using Eq.~(\ref{eq:u}).
FE-PINN solutions are shown to provide a faithful match with FE solutions, consistent with other PINN techniques~\cite{raissiPhysicsinformedNeuralNetworks2019,gaoPhyGeoNetPhysicsinformedGeometryadaptive2021}.
Note that stresses and strains can be computed from the FE-PINN solution using standard FE methods~\cite{hughesFiniteElementMethod2000}.

\subsection{Testing on unseen geometries}

Next we explore the capacity for FE-PINNs to act as surrogate models by testing them against unseen geometries.
We want to reiterate that the vast majority of PINNs cannot be applied to unseen geometries in this way.
We explore the behavior of FE-PINNs when trained on combinations of one or three different geometries.

\subsubsection{Training with a single geometry}

\begin{figure}
    \centering
    \includegraphics[width=0.95\textwidth]{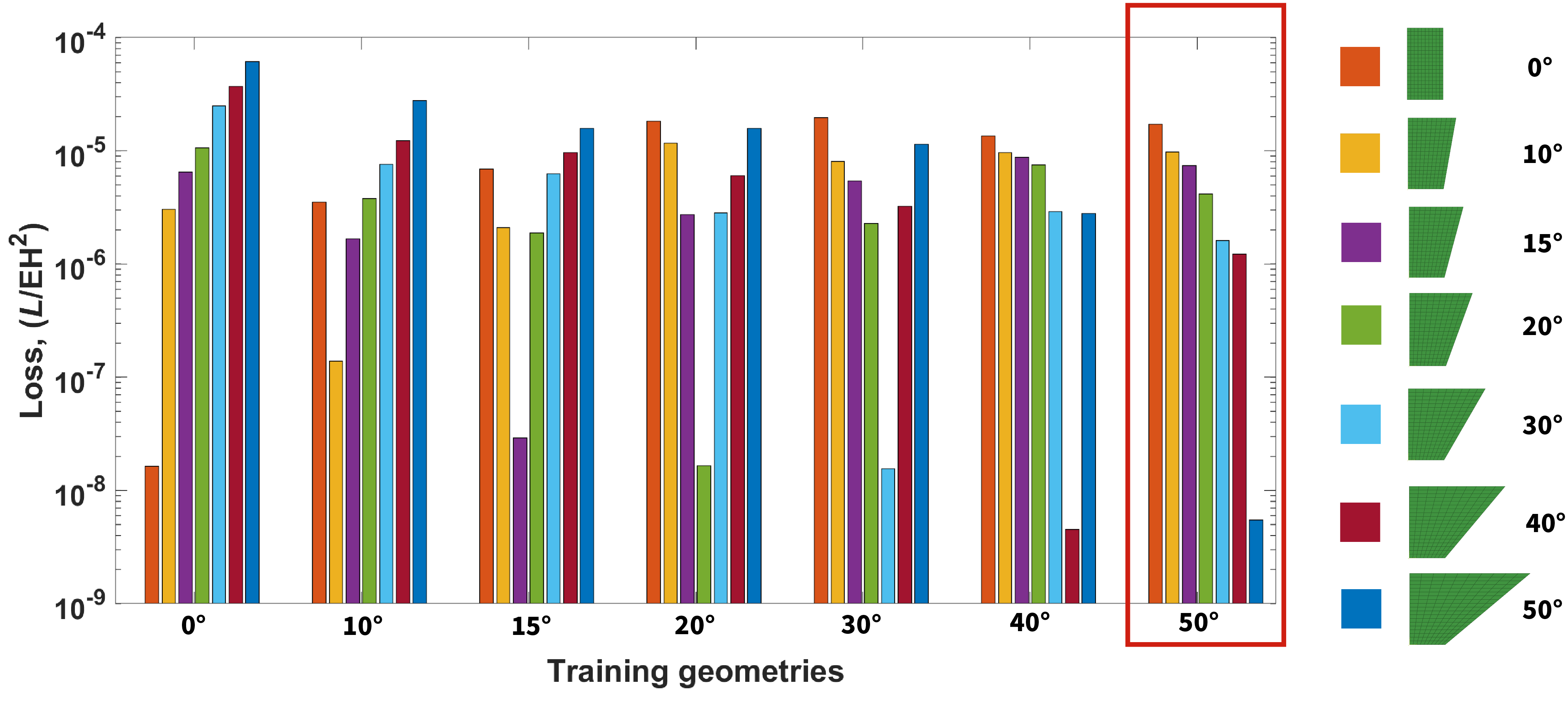} \\
    (a) \\
    \vspace{0.1in}
    \includegraphics[width=0.65\textwidth]{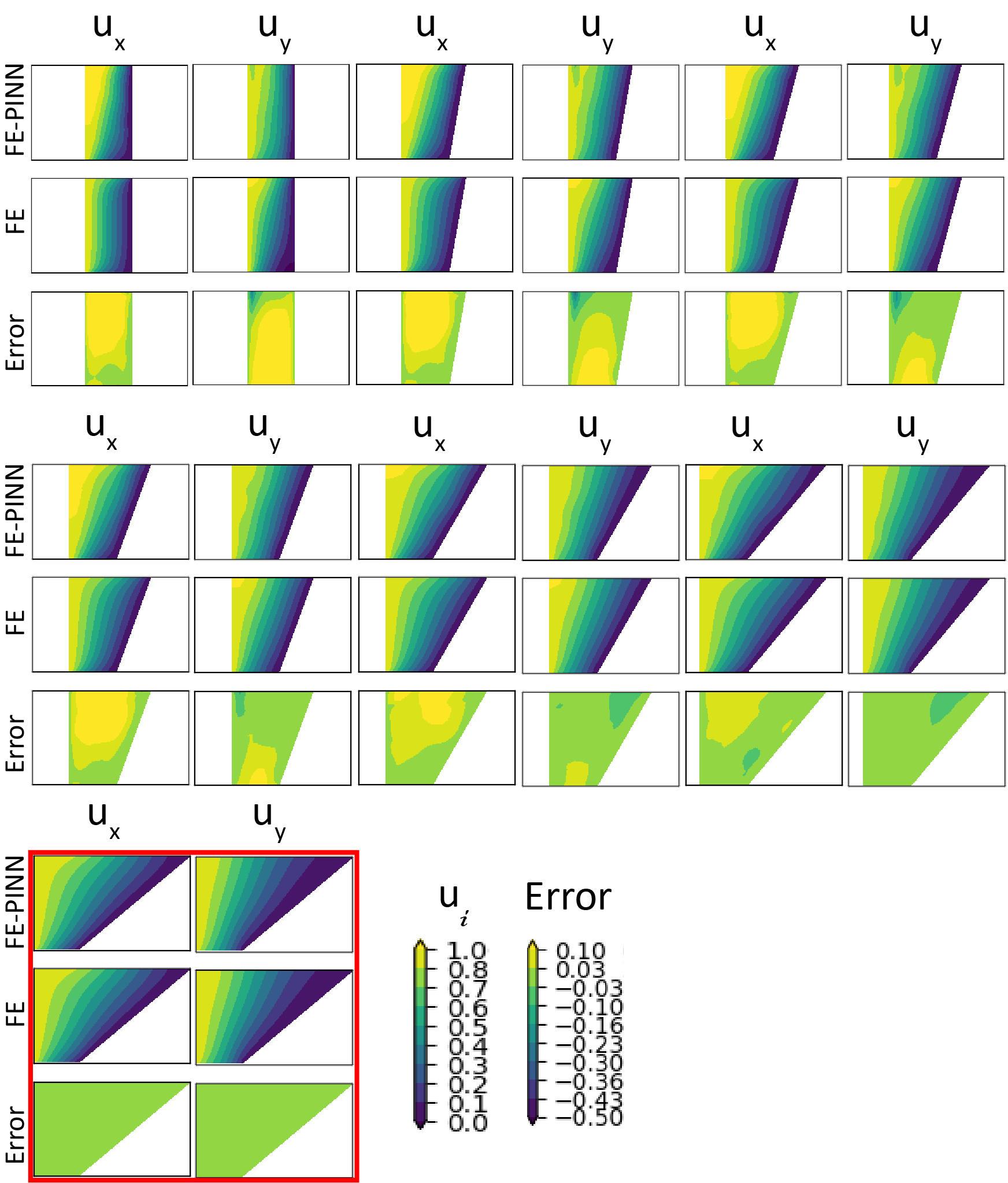} \\
    (b)
    \caption{(a) Losses for all ``wedged block'' geometries using FE-PINNs trained on different single geometries. (b) Displacement field comparison for the best performing FE-PINN. Best performing training case indicated by a red box.}
    \label{fig:r1_wedge}
\end{figure}

Training and testing losses for FE-PINNs trained on single wedged block geometries are shown in Fig.~\ref{fig:r1_wedge}(a).
Losses are grouped based on the geometry used for training, which is indicated in the $x$-axis by the wedge angle.
The figure demonstrates that the training loss is typically about a factor of 100 to 1000 lower than the testing loss.
As expected, geometries that are closer to the training geometry (e.g., similar wedge angle) have lower losses.
Fig.~\ref{fig:r1_wedge}(b) shows the displacement (and error) fields obtained using the FE-PINN with the lowest average loss, which turns out to be the one trained on the $50^\circ$ block, as indicated by the red box.
Examining the displacement and error fields it is clear that the displacement fields for nearby geometries are similar to their FE solutions.
However FE-PINN solutions for geometries with much different wedge angles do not match FE solutions very closely.

\begin{figure}
    \centering
    \includegraphics[width=0.95\textwidth]{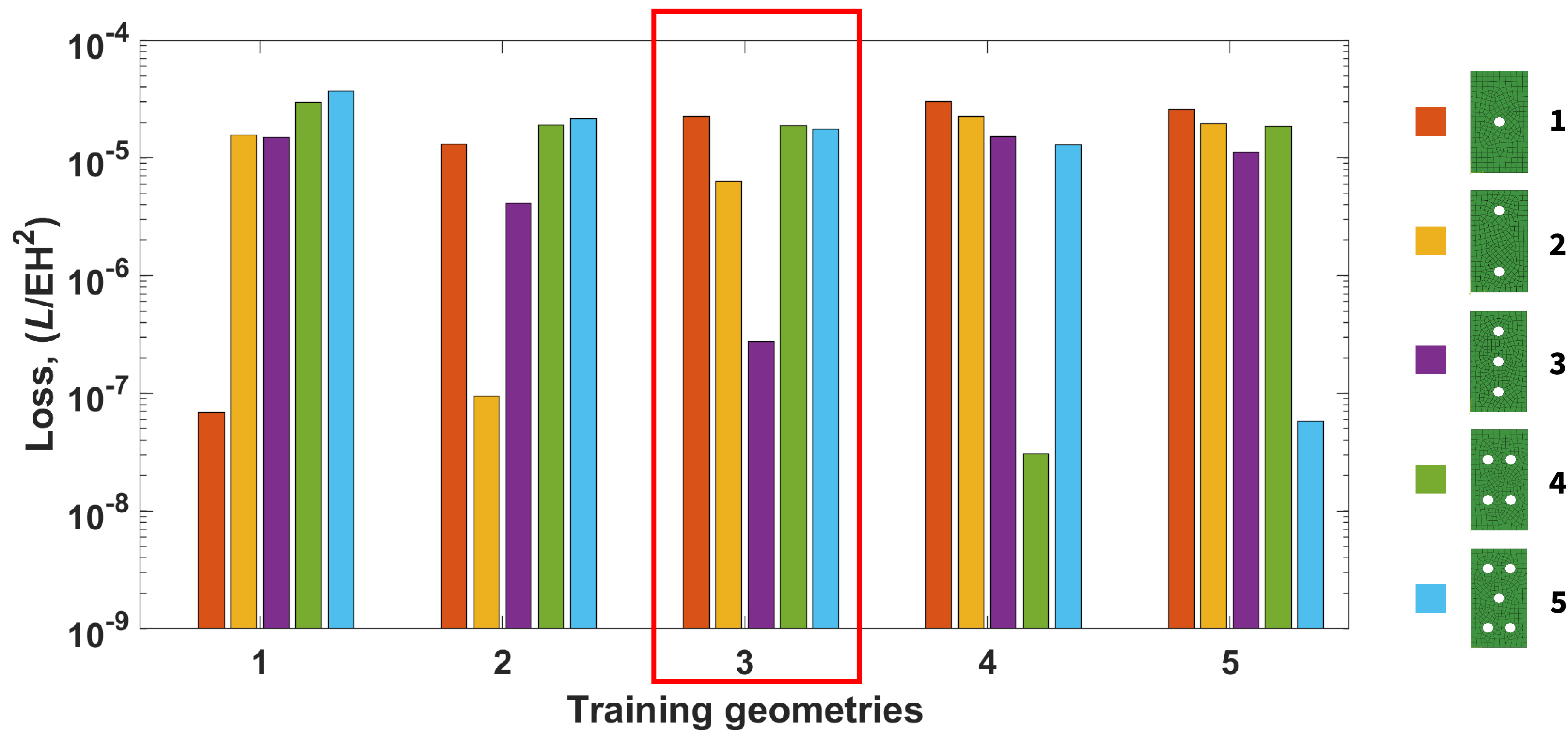} \\
    (a)
    \vspace{0.1in} \\
    \includegraphics[width=0.65\textwidth]{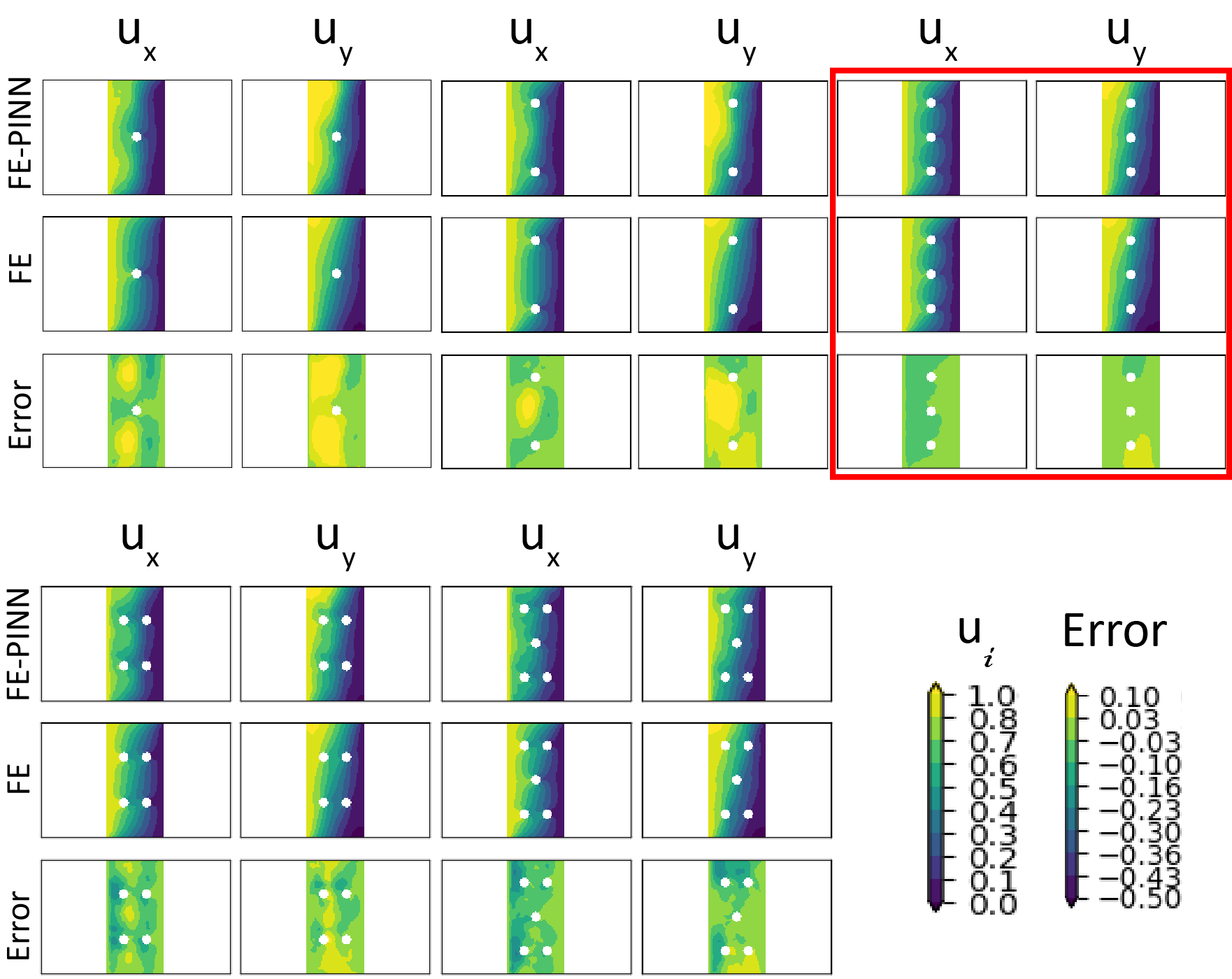}
    \\ (b)
    \caption{(a) Losses for all ``block with hole'' geometries using FE-PINNs trained on different single geometries. (b) Displacement field comparison for the best performing FE-PINN. Best performing training case indicated by a red box.}
    \label{fig:r1_hole}
\end{figure}

Fig.~\ref{fig:r1_hole} shows results for the hole geometries.
In this case, the training geometery is indicated in the $x$-axis based on the number of holes.
Results with these training geometries are similar to the wedged block results, however since each hole geometry is distinct (e.g., adding a new hole is major change in the geometry) the testing losses are generally high; there are no ``nearby geometries'' which the NN is able to extrapolate to with moderate accuracy.

\subsubsection{Training with three geometries}

The results above indicate that training on a single model will not yield a FE-PINN with general predictive capacity.
This is not a surprise since NNs are fundamentally interpolation engines, and interpolation is impossible when only trained on a single geometry.
Here we explore training FE-PINNs using all possible combinations of three geometries.

\begin{figure}
    \centering
    \includegraphics[width=0.65\textwidth]{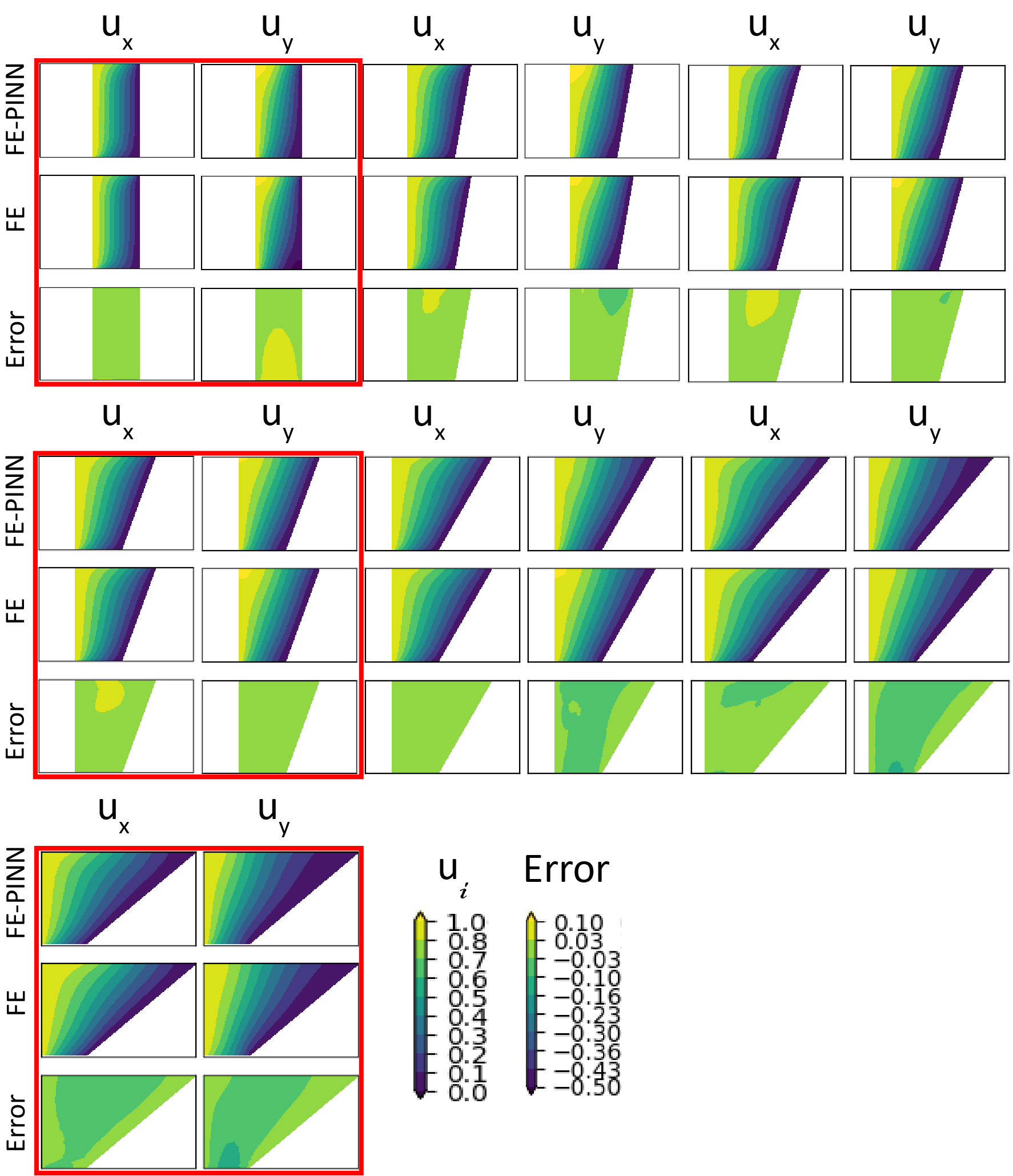} \\
    (c)
    \caption{(a,b) Losses for all ``wedged block'' geometries using FE-PINNs trained on different combinations of three geometries. (c) Displacement field comparison for the best performing FE-PINN. Best performing training case indicated by red boxes.}
    \label{fig:r3_wedge}
\end{figure}

Fig.~\ref{fig:r3_wedge} presents the results for the wedged block geometries.
The number of possible geometry combinations is 35, so the final loss figure is split in two; Fig.~\ref{fig:r3_wedge}(a) shows results for all training cases where the $0^\circ$ block is included in the training set and Fig.~\ref{fig:r3_wedge}(b) shows all other cases.
The first observation to make across all training cases is that the testing losses are lower in general when trained on three geometries.
For example, essentially all losses in Fig.~\ref{fig:r3_wedge} are below $10^{-5}$, which is not the case in Fig.~\ref{fig:r1_wedge}.
Secondly, the training losses are comparable to those obtained during single geometry training runs, on the order of $10^{-8}$.
This indicates that the FE-PINNs are able to learn multiple geometries at the same time without compromising training convergence.
In terms of testing losses, we see the same trend as with single training geometries that testing losses are lower for ``nearby'' geometries.
As a result, the best performer (indicated by the red box) features wedge angles spread over the geometric space ($0^\circ$, $20^\circ$, and $50^\circ$).
Losses for all geometries with this model are below $10^{-6}$.
Displacement fields for this FE-PINN are shown in Fig.~\ref{fig:r3_wedge}(c).
In contrast to Fig.~\ref{fig:r1_wedge}(b), displacement fields are relatively faithful to the FE solution across the whole geometric space.

\begin{figure}
    \centering
    \includegraphics[width=1.0\textwidth]{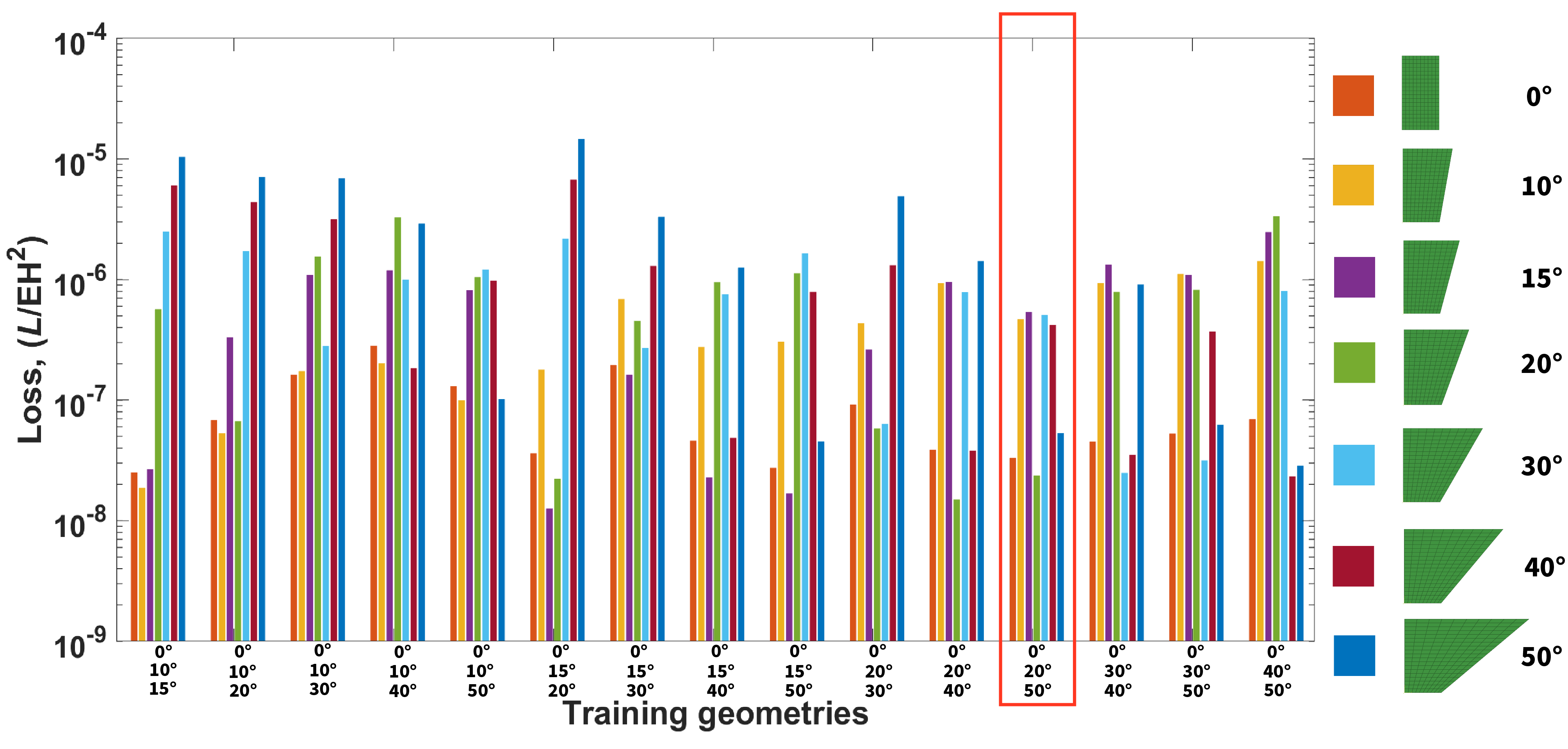} \\
    (a) \\
    \includegraphics[width=1.0\textwidth]{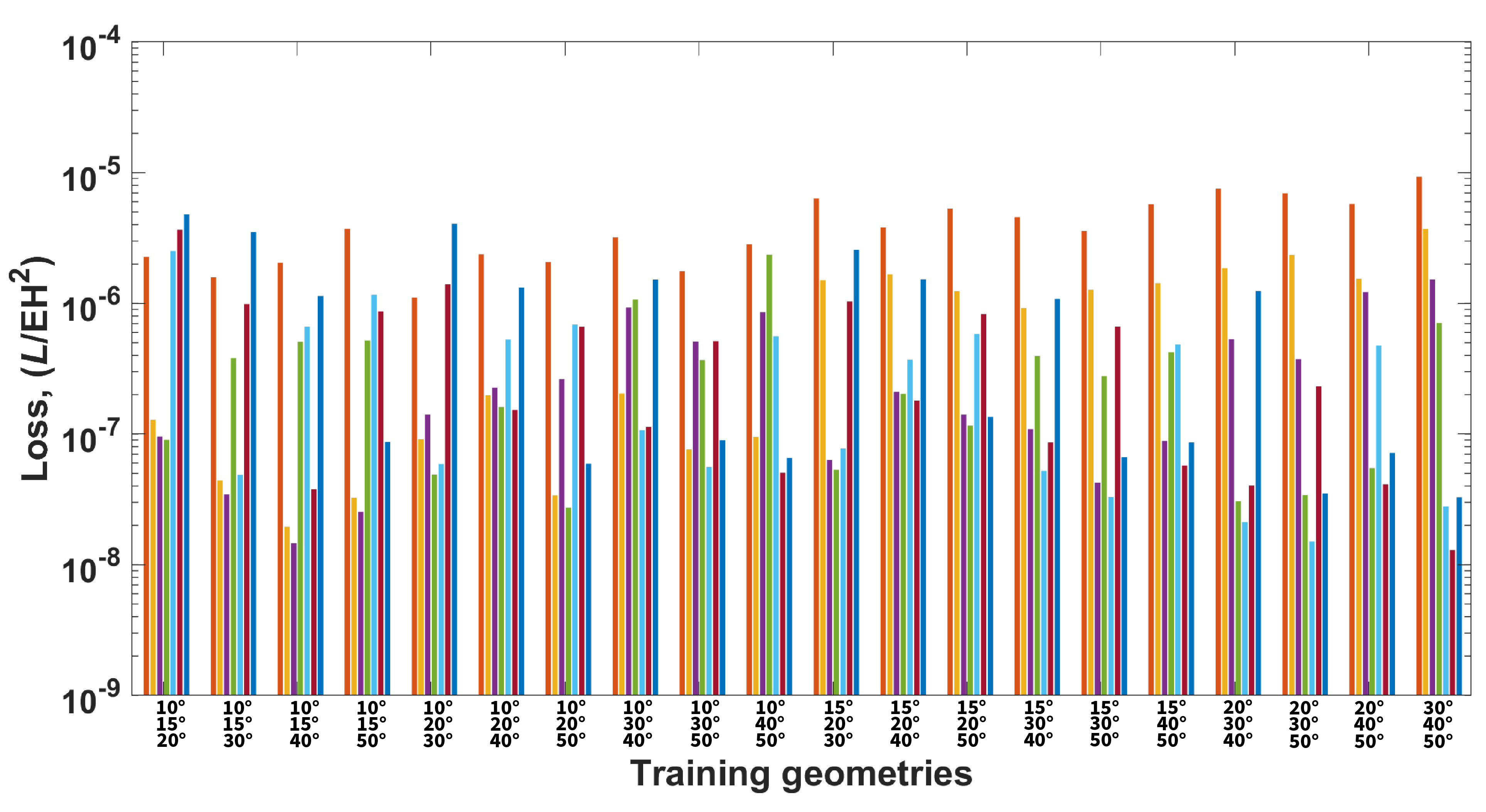} \\
    (b) 
\end{figure}

\begin{figure}
    \centering
    \includegraphics[width=1.0\textwidth]{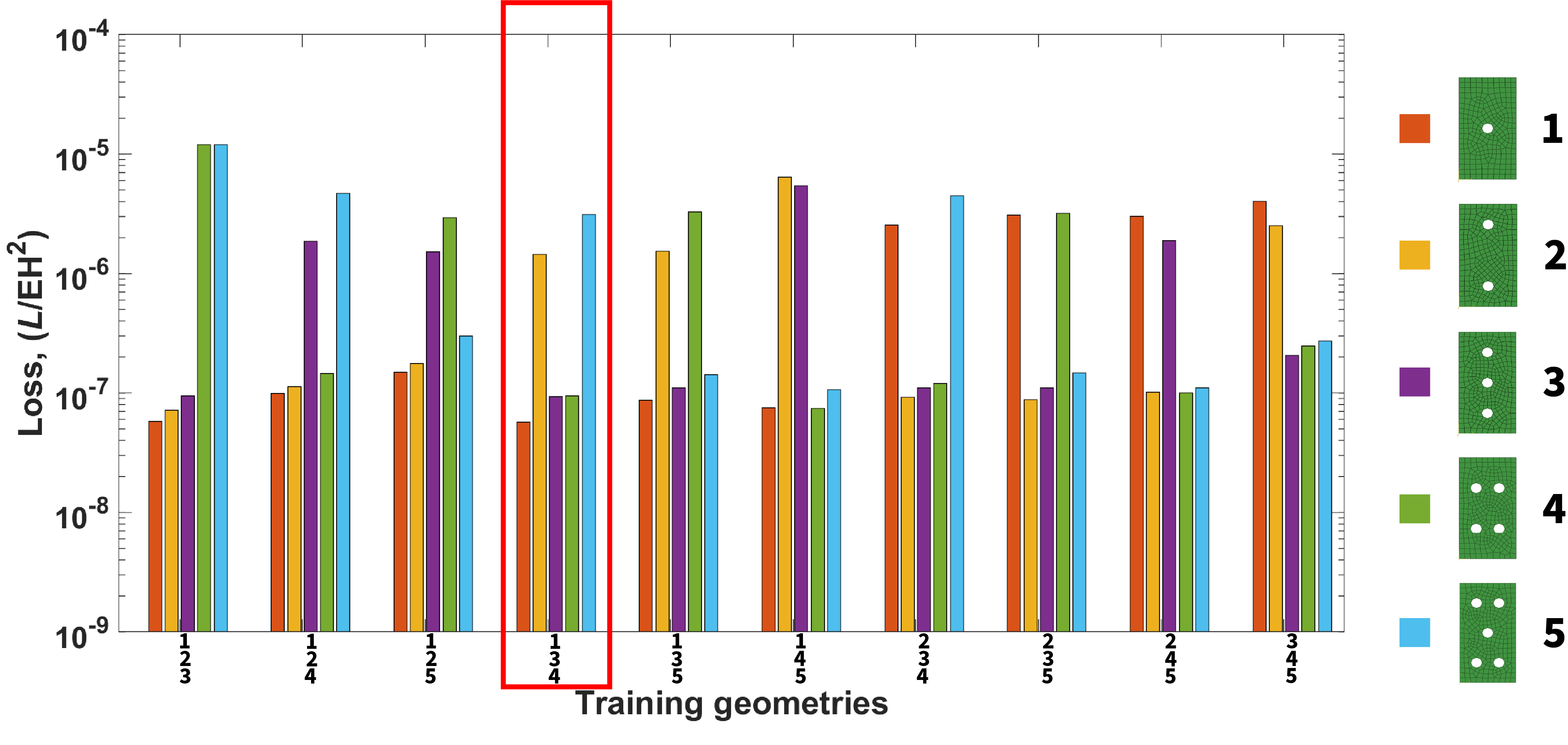} \\
    (a) \\
    \vspace{0.1in}
    \includegraphics[width=0.65\textwidth]{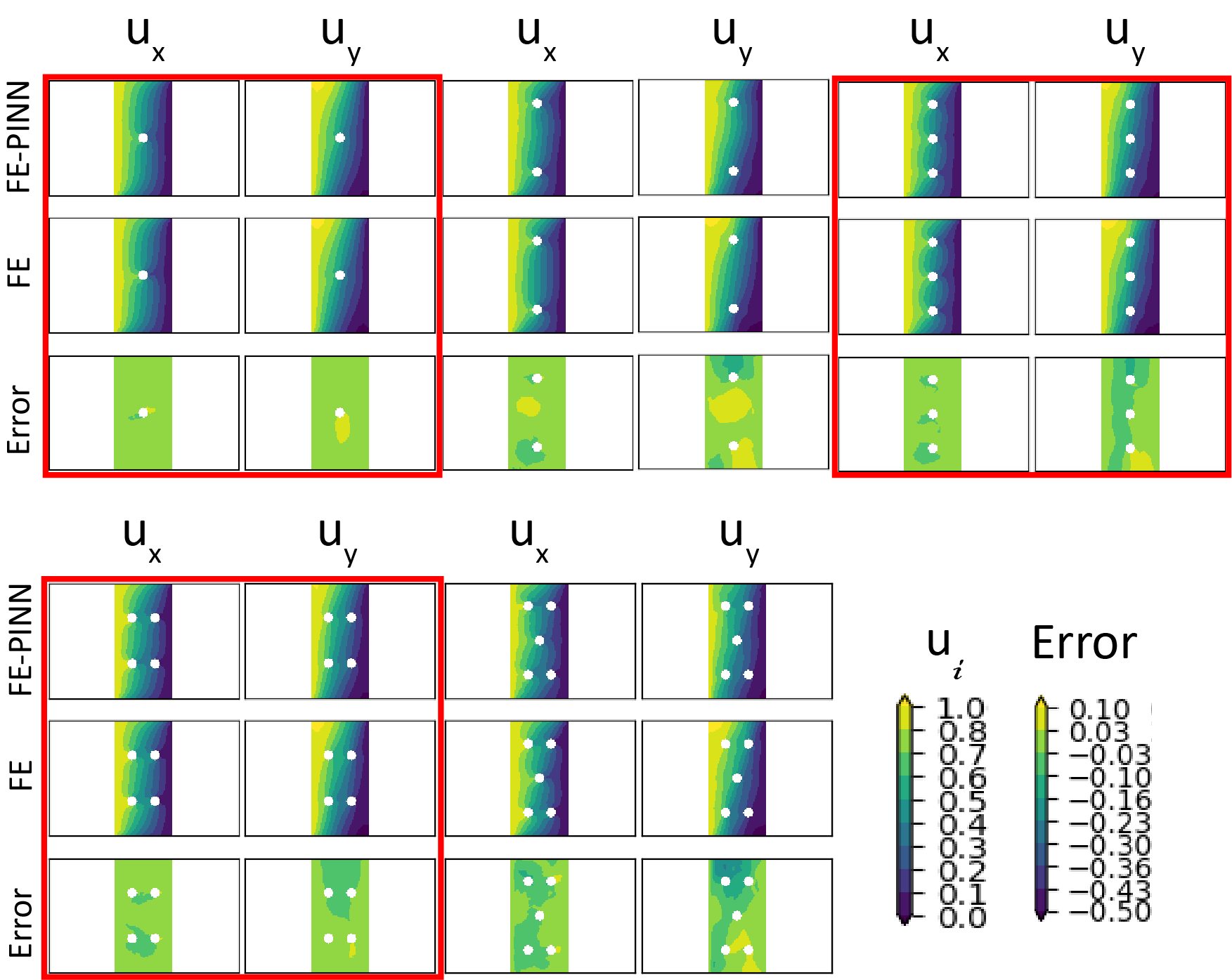} \\
    (b)
    \caption{(a) Losses for all ``block with hole'' geometries using FE-PINNs trained on different combinations of three geometries. (b) Displacement field comparison for the best performing FE-PINN. Best performing training case indicated by red boxes.}
    \label{fig:r3_hole}
\end{figure}

Fig.~\ref{fig:r3_hole} presents results for the blocks with holes when trained on three geometries.
Once again, the testing losses are generally lower (below $10^{-5}$) in comparison to the single training geometry results in Fig.~\ref{fig:r1_hole}, while the training losses are comparable.
This demonstrates that even when trained on geometries that are very different from one another, the FE-PINN is able to generally improve its predictions.
Unlike the wedged block results, here results are quite similar for all training geometry combinations.
The best performer in this case was trained on the 1, 3, and 4 hole blocks, and associated displacement fields are shown in Fig.~\ref{fig:r3_hole}(b).
Predictive performance (for the 2 and 5 hole blocks) is not quite as strong as it was in the wedged block study.
Again, the key difference here is that the geometry variation is not smoothly graded.

\section{Discussion}
\label{sec:discussion}

Because stencil convolution can be cast in terms of linear tensor operations, it is likely to have comparable computational cost to traditional convolutions, especially when sparse tensors are used to avoid unnecessary multiplications by zero.
However the stencil convolution tensor $S_{klmn}$ needs to be precomputed prior to FE-PINN evaluation.
Hence, to have the overall cost of FE-PINN evaluation as low as possible, an efficient subroutine for constructing $S_{klmn}$ must be employed.
Each element of $S_{klmn}$ can be computed independently, making its construction amenable to GPU (or parallel) computation, which should help to reduce the computation time.
However, a key step in constructing $S_{klmn}$ is evaluating the inverse isoparametric map for the chosen finite elements.
For the test cases considered here with linear 2D elements this mapping is analytical, but for other element types (e.g., 3D elements, quadratic 2D elements) it is not.
Researchers have applied techniques to approximately evaluate the inverse isoparametric map in some cases~\cite{yuanInverseMappingDistortion1994,liEfficientInverseIsoparametric2014}, but this will of course drive up the computational cost of FE-PINN evaluation.

In this work we only explored FE-PINNs featuring convolutional layers with a single stencil spacing $\Delta$ across all layers.
FE-PINNs afford a rather more broad set of NN architectures, however.
Firstly, the stencil tensor $s_{ikl}$ used to perform convolutions is completely arbitrary in the sense that stencil points can be arranged in any way.
We limited our analyses to uniform grids defined via Eq.~(\ref{eq:sikl}), however non-uniform stencil point arrangements may lead to better performance.
Secondly, multiple stencil tensors can be used within the same FE-PINN, making it possible for convolutions at multiple length scales to be leveraged simultaneously.
And thirdly, other classes of NN operators can also be executed using stencils.
For example, pooling and concatenation operations can be executed using stencils.
Together, these features of FE-PINNs provide a large NN architectural space which could be explored in the training of a ``large'' FE-PINN that is trained on 100s or 1000s of FE models with varying conditions.

Despite the fact that our FE-PINNs were trained on FE meshes with varying characteristics (structured vs. unstructured, distorted vs. undistorted, variable connectivity of nodes), we saw no trends indicating mesh sensitivity.
In other words, stencil convolutions do not appear to be significantly influenced by the FE mesh used during training or testing.
In contrast, we argue that graph convolutions are more likely to be influenced by the mesh, since the convolutions are ``slaved'' to it.
All results presented above used meshes with similar element sizes and the influence of variable element size was not explored.
In general, one should train a FE-PINN using a suitably refined mesh to approximate the solution of the weak form to the desired degree of accuracy; any numerical error from the FE solution will be ``baked into'' the FE-PINN.
Users may also want to change the element size after training.
Performance of FE-PINNs with variable element sizes has yet to be explored.
The neural operator approach developed by Kovachki et al.~\cite{kovachkiNeuralOperatorLearning2023} has the advantage of being discretization invariant, meaning that the NN solution converges as the grid/mesh is refined.
We may expect FE-PINNs to also exhibit this property since, like neural operators, the inputs and outputs of a FE-PINN are functions (defined via Eq.~(\ref{eq:u})).
Further research is necessary to explore the discretization sensitivity of FE-PINNs.

One interesting observation from the wedged block results in Fig.~\ref{fig:r3_wedge} is that all FE-PINNs which are not trained on the $0^\circ$ block exhibit large testing error on the $0^\circ$ block.
This is true even if the ``nearby'' $10^\circ$ block is included in the training set, even though in all other cases geometries test well if one of their nearby geometries is used for training.
This behavior has two possible explanations.
One is that the $0^\circ$ block has an undistorted mesh while all other wedged blocks have distorted meshes.
To determine if the mesh was the cause, an additional set of results was generated using an unstructured mesh for the $0^\circ$ block, yielding very similar results (not shown).
This observation further strengthens the finding that stencil convolutions are mesh insensitive.
The other possible explanation is that the $0^\circ$ block is too ``perfect'' in comparison to the other blocks, making it challenging for a FE-PINN to perform well when only trained on ``imperfect'' geometries with a non-zero wedge angle.
This result indicates that care must be taken when selecting a training set for FE-PINNs.

\begin{figure}
    \centering
    \begin{tabular}{cc}
        \includegraphics[width=0.47\textwidth]{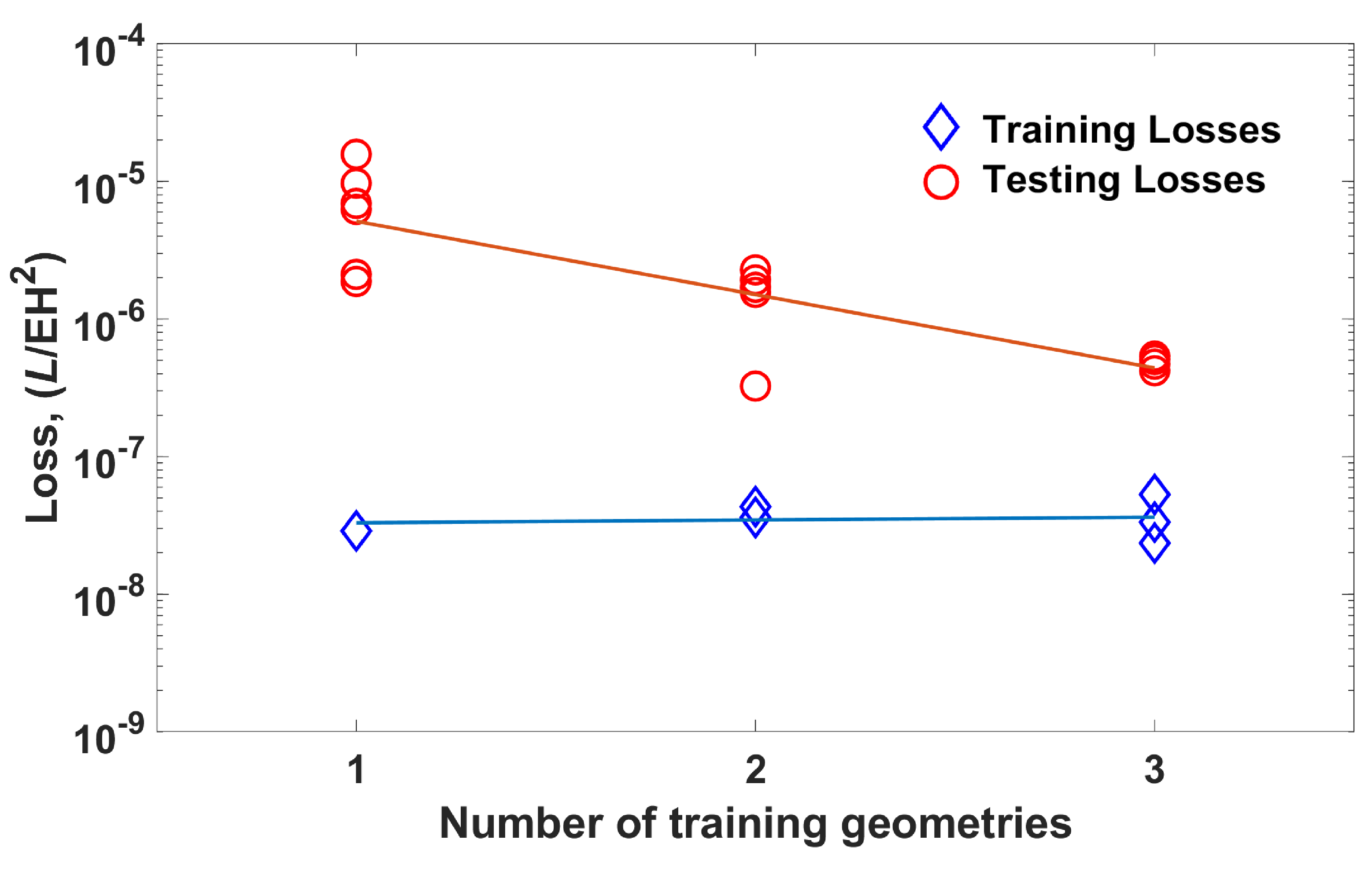} & 
        \includegraphics[width=0.47\textwidth]{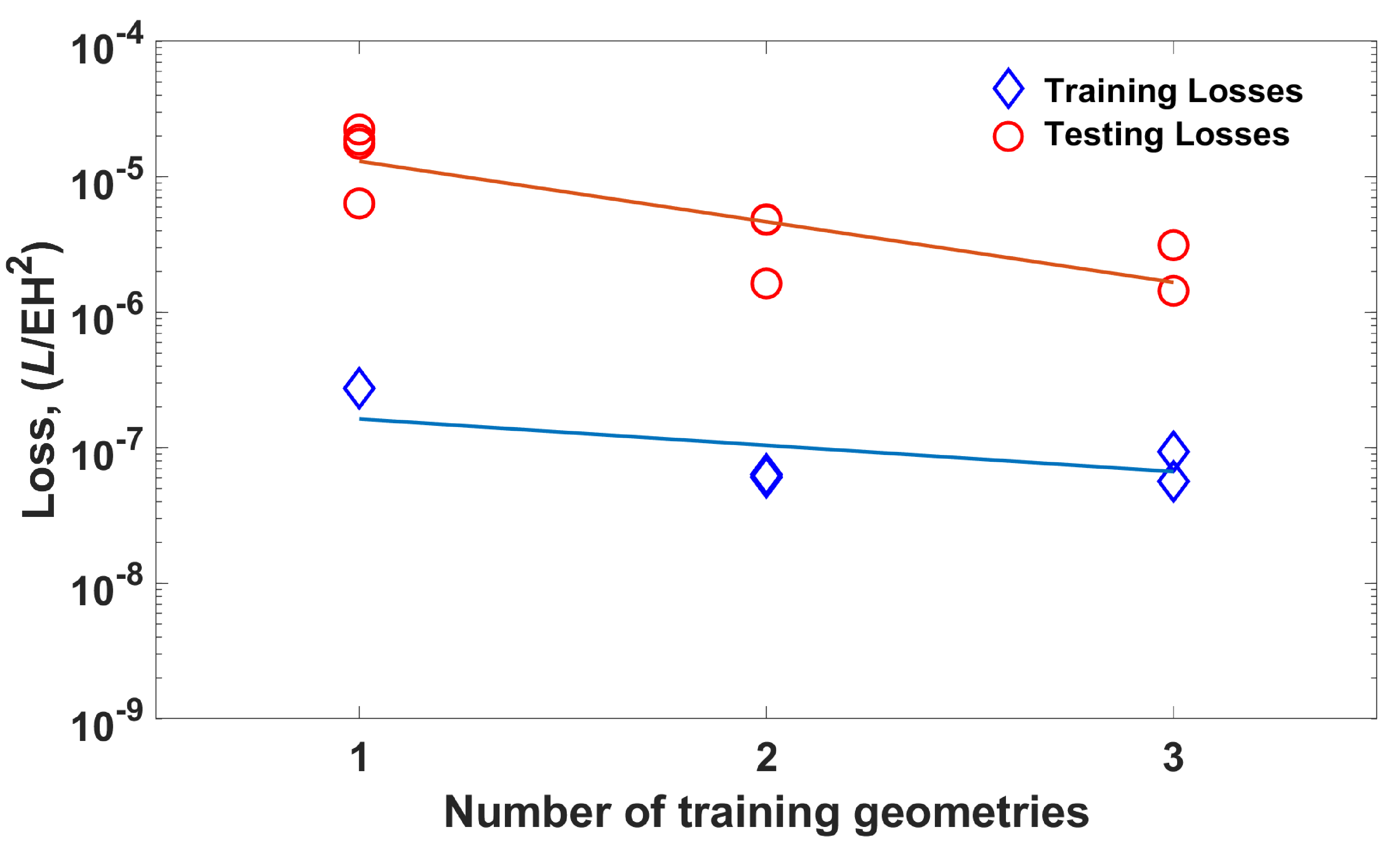} \\
        (a) & (b) 
    \end{tabular}    
    \caption{Training and testing losses from best performing FE-PINNs as a function of number of training geometries for (a) wedged block and (b) block with holes geometries. Lines are linear regression fits.}
    \label{fig:loss_summary}
\end{figure}

One key question is whether PINNs truly ``learn'' the physics underlying the governing questions on which they are trained.
If they do learn in this way, we would expect them to have stronger predictive capacity than if they just memorized a handful of solutions which they interpolate among.
Our results provide some evidence that our FE-PINNs seem to be generalizing their predictions, which may indicate a deeper connection to the physics.
For example, in the $u_x$ fields for the holes with blocks, the main influence of the holes is to locally amplify the displacements.
This is challenging for the FE-PINN to capture during testing since the holes in the blocks vary in their positions.
Nonetheless, the $u_x$ fields in Fig.~\ref{fig:r3_hole}(b) show that the FE-PINN captures the local  amplification of $u_x$, at least semi-quantitatively.
Additional evidence can be found by looking at the losses of the best performing FE-PINNs as a function of number of training geometries.
Fig.~\ref{fig:loss_summary} shows these results for (a) wedged block and (b) block with hole geometries.
We also present in this figure results for FE-PINNs trained on two geometries, which were not presented in the results.
This figure clearly shows that as the number of training geometries increases the testing losses decrease while the training losses stay comparatively flat.
This demonstrates that the FE-PINN gets better at generalizing as it learns more physics (in the form of novel geometries).

Above we only explored the performance of FE-PINNs on 2D, linear problems with varying geometry.
In principal, FE-PINNs can be applied to nonlinear, 3D problems with varying geometry, BCs, properties, body forces, etc.
The extension to 3D should be possible by defining a 3D stencil tensor and then applying the same tensor operations (with one additional index for the convolutional weights and stencil points).
Extending FE-PINNs to problems with more than just the geometry varying is also expected to be possible by constructing  additional input fields (defined via nodal values) to specify each variable model parameter, as is done in PhyGeoNet~\cite{gaoPhyGeoNetPhysicsinformedGeometryadaptive2021}.
These additional input fields are then additional input channels at the input layer of the NN.
As long as the input fields are rich enough to specify the changes in the problem setup, the NN should be able to establish a mapping between them and the associated solution.
Such extensions are actively under development by the authors.
Extension to nonlinear problems is more challenging because it requires communication between the NN training code and FE software package to evaluate the loss function, Eq.~(\ref{eq:loss}).
Furthermore, the backward associated with the loss function needs to be computed by the FE software.
These methods are also under active development.

\section{Conclusions}
\label{sec:conclusions}

We demonstrated a new class of PINNs which leverages the FE method to perform convolutions, compute the physics-based loss, and enforce the BCs.
A new class of convolutions was introduced that utilizes a stencil tensor to define a set of stencil points where fields are evaluated through the inverse isoparametric map of the FE method.
The resulting FE-PINNs were trained on a variety of scenarios considering deformation of 2D, linear elastic blocks with varying geometry.
Robust training convergence was observed and the influence of the stencil tensor was explored, exhibiting an ideal range of length scales (i.e., values of $\Delta$) for convolutions to be performed that minimizes the training loss.
Training FE-PINNs on multiple different geometries enables systematically better testing predictions on unseen geometries without sacrificing training convergence.
The FE-PINN method is extensible to nonlinear, 3D problems with variable geometry, BCs, and properties, providing a general purpose surrogate modeling framework.

\section*{Acknowledgements}
The authors thank Dr. Manish Vasoya for feedback on the manuscript.
This material is based upon work supported by the National Science Foundation under Grant No. 2237039. 
Research to initiate development of the method was sponsored by the Army Research Office, United States and was accomplished under Grant Number W911NF-21-1- 0086. The views and conclusions contained in this document are those of the authors and should not be interpreted as representing the official policies, either expressed or implied, of the Army Research Office or the U.S. Government. The U.S. Government is authorized to reproduce and distribute reprints for Government purposes notwithstanding any copyright notation herein.

\section*{Conflict of Interest Statement}
The authors have no conflicts to disclose.

\section*{Author Contributions}
\textbf{Pranav Sunil}: Data curation, Formal analysis, Investigation, Methodology, Software, Supervision, Validation, Visualization, Writing – review and editing
\textbf{Ryan Sills}: Conceptualization, Formal analysis, Funding acquisition, Investigation, Methodology, Project administration, Software, Validation, Writing – original draft, Writing – review and editing

\section*{Data Availability Statement}
Code used to produce results is available in the following public GitLab repository: https://gitlab.com/mmod\_public/fepinn. We also provide in this repository neural networks (.pth files) used to produce results in Figs.~\ref{fig:r1_wedge}(b), \ref{fig:r1_hole}(b), \ref{fig:r3_wedge}(c), and \ref{fig:r3_hole}(b), along with all files necessary to train new neural networks.



  \bibliographystyle{elsarticle-num} 
  \bibliography{reference}





\end{document}